\documentclass[12pt]{iopart}
\usepackage{iopams}  

\usepackage{graphicx}  
\usepackage{dcolumn}   
\usepackage{bm}        
\usepackage{amssymb}   

\begin{document}

\newcommand{\EuYSO}{Eu$^{3+}$:Y$_2$SiO$_5$~}
\newcommand{\Eu}{Eu$^{3+}$}
\newcommand{\YSO}{Y$_2$SiO$_5$}
\newcommand{\F}{$^7$F$_0$}
\newcommand{\D}{$^5$D$_0$}
\newcommand{\xition}{\F$\rightarrow$\D}

\title{Shifts of optical frequency references based on spectral-hole burning in \EuYSO}

\author{Michael~J.~Thorpe\footnote{Present address: Bridger Photonics, 2310 University Way, Bozeman, MT 59715, USA}, David~R.~Leibrandt, and Till~Rosenband}
\address{National Institute of Standards and Technology, 325 Broadway St., Boulder, CO 80305, USA}
\ead{till.rosenband@nist.gov}

\date{\today}

\begin{abstract}
Several properties of \EuYSO spectral holes are measured, to assess the suitability of broad-band hole-patterns for use as laser-frequency references.  We measure frequency shifts due to magnetic fields, side-features of neighboring spectral holes, and changing optical probe power. A precise calibration of a temperature insensitive point is also performed, where the temperature-induced frequency shift is canceled to first order by the pressure-induced shift from the crystal's helium-gas environment.  
\end{abstract}

\pacs{42.62.Fi, 78.47.nd, 32.60.+I}
\maketitle

\section{Introduction}

Frequency-stable laser-local-oscillators (LLO) are critical components of the new generation of optical atomic clocks \cite{Chou2010a,Jiang2011,Kessler2012}. The stability of such clocks is presently limited by thermomechanical noise in the cavity-stabilized LLOs \cite{Numata2004,Notcutt2006}. Quieter LLOs would enable optical clocks with improved stability, yielding faster comparison measurements among different clocks and more precise searches for the variation of fundamental constants \cite{Rosenband2008,Blatt2008,Chou2010b}. Quieter LLOs may also yield lower-phase-noise microwave oscillators by frequency-division via femtosecond laser frequency combs \cite{Zhang2010,Fortier2011,Benedick2012}.

Spectral-hole-burning (SHB) laser frequency stabilization~\cite{Sellin1999a,Strickland2000,Bottger2003,Julsgaard2007} in \Eu-doped \YSO~has recently approached the performance of state-of-the-art cavity-stabilized LLOs~\cite{Thorpe2011} and indirect measurements suggest that the fundamental noise floor may be significantly lower.  Because of the long lifetime of spectral holes (of order $10^6$~s at 3.7~K \cite{Konz2003}), the large number of dopant ions within a crystal (of order $10^{18}$/cm$^3$), and the narrow linewidth of each absorber, such crystals can serve as precise memories for laser frequency.  Comparison of the current laser frequency with the frequencies imprinted in the memory enables frequency stabilization.  

Inside the crystal, the narrow \xition~optical transition of each \Eu~ion is shifted by the local crystal field, which is slightly random, such that each \Eu~dopant has a fixed transition frequency within the typical 1-30 GHz inhomogeneous linewidth.  The isolated nature of rare-earth ion inner-shell transitions has allowed dephasing rates below 500~Hz to be observed in this type of crystal \cite{Yano1991,Macfarlane1994,Konz2003}. In \Eu, the \F~ground-state comprises several long-lived hyperfine levels (nuclear spin I=5/2). The hyperfine level splittings are of order 100 MHz, and for any given laser frequency within the inhomogeneous linewidth, particular combinations of crystal-fields and hyperfine levels are resonant.  The resonant population of \Eu~ions is excited by the laser, and a significant fraction decays into different hyperfine levels, causing a transparency, or spectral-hole, at the laser frequency.

High-resolution spectroscopy has shown that the environmental sensitivity of \EuYSO spectral holes to temperature, pressure, and accelerations are all smaller than that of Fabry-P\'{e}rot cavities \cite{Thorpe2011}, and frequency drift rates can be lower than those of cavity-stabilized lasers \cite{Chen2011}.  These results suggest that spectral holes in \EuYSO could form a frequency-stable flywheel for optical atomic clocks and potentially extend the coherence time of atom-laser interactions.

Unlike Fabry-P\'{e}rot cavities, whose oscillation period is directly proportional to an artificial length, the optical frequency of spectral holes depends primarily on the atomic Hamiltonian of dopant ions. Spectral holes are less sensitive to mechanical perturbations, which couple only weakly through the field of the host crystal, but may be more sensitive to ambient electric and magnetic fields.  Moreover, probing of the spectral holes causes additional burning that can modify the hole shape and frequency.  This work aims to determine the extent to which such frequency shifts will limit the performance of SHB LLOs, and we describe measurements of \EuYSO spectral-hole properties including the magnetic field sensitivity, perturbations to the spectral-hole pattern due to side-holes and anti-holes, and frequency shifts due to fluctuations in the optical probe power.  Temperature-dependent shifts are also measured, and it is shown that the linear dependence of the hole frequency on temperature can be eliminated by placing the crystal inside a sealed chamber that is filled with a particular pressure of helium gas.  With the current level of environmental stability, fluctuations in these frequency shifts are small enough to allow for laser stabilization with a fractional-frequency noise floor below $1 \times 10^{-17}$.  This would represent an order of magnitude improvement over state-of-the-art laser frequency stabilization using Fabry-P\'{e}rot cavities \cite{Jiang2011,Kessler2012}.

This paper proceeds as follows: Section~\ref{sec:setup} describes the experimental setup.  Measurements of \EuYSO spectral-hole properties are presented in Sections~\ref{sec:magnetic} (magnetic field sensitivity) and \ref{sec:tempAndPower} (temperature and optical probe power sensitivity).  Perturbations to spectral-hole patterns due to side-holes and anti-holes are discussed in Section~\ref{sec:sideHoles}.  A summary of the frequency shifts of optical frequency references based on spectral-hole burning in \EuYSO~is provided in Section~\ref{sec:shifts}.

\section{Setup}\label{sec:setup}

\begin{figure}
\begin{indented}
\item[]\includegraphics[width=0.6\columnwidth]{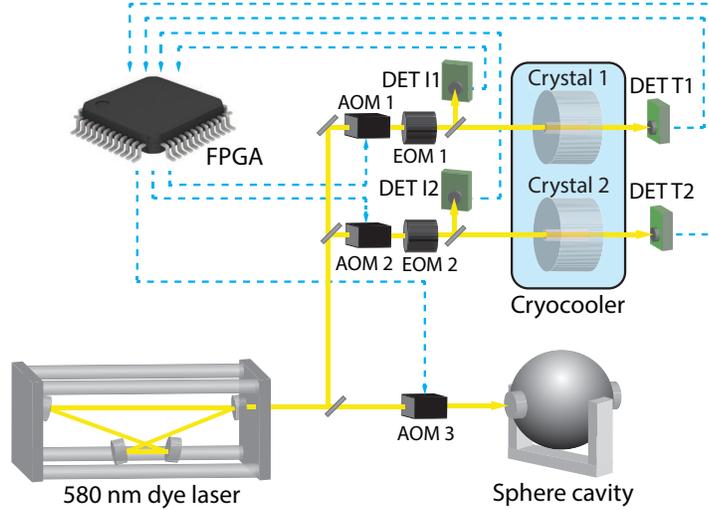}
\end{indented}
\caption{\label{fig1} Experimental setup for measuring the properties of \EuYSO spectral holes.  A 580~nm dye laser is pre-stabilized to a spherical Fabry-P\'{e}rot reference cavity. The laser burns and probes spectral holes in two \EuYSO crystals located in a single closed cycle cryostat.  Spectroscopy is controlled by a field-programmable gate array (FPGA).  AOM: acousto-optic modulator, EOM: electro-optic modulator, DET: detector.}
\end{figure}

A schematic of the experimental setup is shown in Figure~\ref{fig1}.  Spectral holes in cryogenically cooled \EuYSO are burned and probed by a dye laser that is pre-stabilized to a spherical Fabry-P\'{e}rot reference cavity \cite{Drever1983, Leibrandt2011a}.  The laser frequency can be tuned coarsely by stepping among longitudinal modes of the reference cavity (3~GHz free spectral range) and finely by shifting the frequency of an acousto-optic modulator (AOM3) located between the laser and the cavity.  The pre-stabilized laser has a fractional frequency stability of $1.2 \times 10^{-15}$ between 0.5~s and 12~s and a typical drift rate of $10^{-14}$/s.  \EuYSO exhibits two inhomogeneously broadened absorption features at 580.039~nm (site 1) and 580.211~nm (site 2) that result from different crystal field shifts at the two distinct positions within the \YSO~unit cell where \Eu~can substitute for Y$^{3+}$; in this work we focus primarily on site 1 spectral holes because they display a smaller temperature shift.

All of the measurements presented in Section~\ref{sec:tempAndPower} were performed on a matching pair of cylindrical samples of 1.0~atomic~(at.)~\%~\EuYSO with a diameter of 10.0~mm and a length of 4.9~mm.  For each sample the laser propagates parallel to the axis of the cylinder (b-axis of the crystal \cite{Li1992}), whose ends are wedged at a 2$^\circ$ angle to prevent etalon effects, and is polarized along the D1 crystal axis.  The samples are cooled to 3.5 to 8.8~K in a closed-cycle cryostat that allows optical access.  Inside the cryostat, the two crystals are enclosed in a single hermetically sealed chamber that is filled with helium gas to provide independent control of temperature and pressure.  The pressure is controlled by a thin tube that connects the sealed chamber to a room-temperature gas manifold and each crystal is mounted in a way that reduces the sensitivity to vibrations \cite{Nazarova2006}.  The measurements presented in Secs.~\ref{sec:magnetic} and \ref{sec:sideHoles} were made with samples of 0.1 and 0.5~at.~\%~\EuYSO in an optical flow cryostat described previously \cite{Thorpe2011}.  All of the crystals used in this work were grown with a natural abundance of europium isotopes (47.8\% $^{151}$Eu and 52.2\% $^{153}$Eu).

Spectroscopy of \EuYSO spectral holes is controlled by a field-programmable gate array (FPGA) that includes an embedded microprocessor.  The laser beams (5~mm diameter) incident on each crystal can be tuned independently over a 600~MHz frequency range and 0.1 to 20~$\mu$W/cm$^2$ intensity by adjusting the radio-frequency (RF) drives of AOM~1 and 2.  Phase modulation sidebands are imprinted on the lasers for Pound-Drever-Hall (PDH) laser frequency stabilization \cite{Julsgaard2007} by electro-optic modulators.  Detectors I1 and I2 measure the power incident on the crystals, which is stabilized at the desired level by feedback to the amplitudes of the RF drives of AOM~1 and AOM~2.  Detectors~T1 and T2 measure the transmission of the laser through crystal~1 and crystal~2, respectively.  The temperature-sensitivity measurement is based on locking the laser frequency to a pattern of several spectral holes by the PDH method, and observing frequency changes with respect to the Fabry-P\'{e}rot reference cavity.  Measurements of magnetic field effects and side-features rely on fitting Lorentzian line-shapes to the measured transmission spectrum of a single spectral hole.

\section{Magnetic field effects}\label{sec:magnetic}

The spectral-hole-burning mechanism in \EuYSO is based on population redistribution among long-lived hyperfine levels, and changes in the magnetic field environment may modify the frequencies and lineshapes of spectral holes due to the Zeeman effect.  However, the \xition~transition of \Eu~has small magnetic moments in both the ground and excited states \cite{Shelby1981}, because only the nuclear angular-momentum is non-zero.

\begin{figure}
\begin{indented}
\item[]\includegraphics[width=0.6\columnwidth]{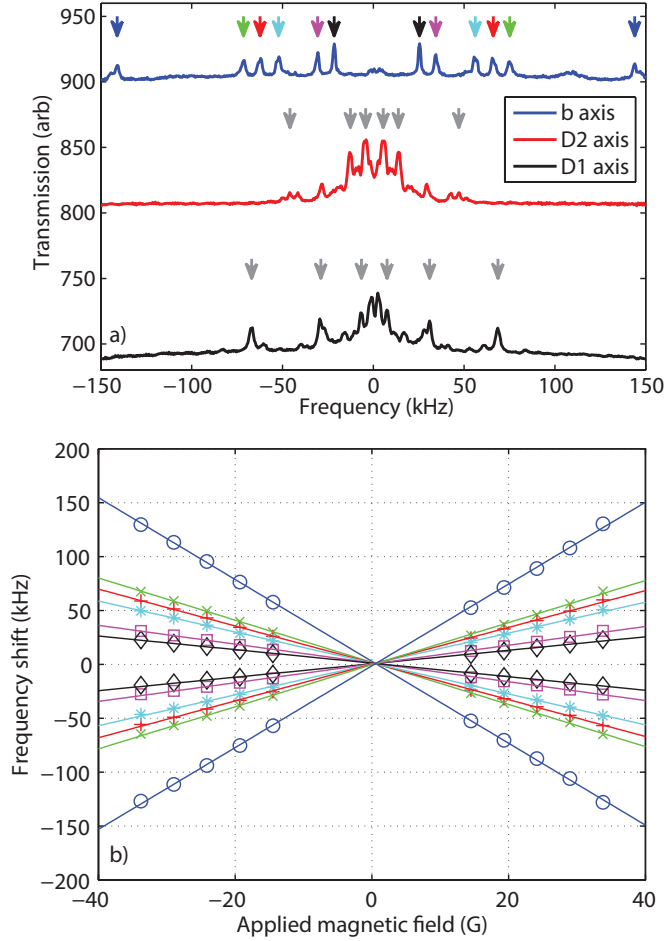}
\end{indented}
\caption{\label{fig2} The linear Zeeman effect in 0.1~at.~\%~\EuYSO site 1 spectral holes at 4.5~K.  (a) A magnetic field applied to spectral holes after they have been written splits the spectral hole into a symmetric set of spectral components.  In this example a spectral hole is written in an applied magnetic field of 0~G and measured in an applied magnetic field of 37~G along each of the crystal axes.  The three measurements are offset vertically for clarity.  (b) Frequency shift of the most prominent spectral components (indicated by vertical arrows in Fig.~\ref{fig2}a) as a function of applied magnetic field along the crystal b axis.}
\end{figure}

Figure~\ref{fig2}a shows the transmission spectrum of a spectral hole written using a 200~ms burn pulse with an intensity of 6.6~$\mu$W/cm$^2$ at zero applied field and subsequently measured in a magnetic field of 37~G~(1~G = 0.1~mT) that is sequentially applied along each of the three crystal axes.  Here the spectral hole, which was a single 1.1~kHz FWHM peak at zero applied field, splits into a symmetric set of transmission peaks corresponding to transitions between pairs of Kramer's doublets.  The Zeeman shifts are different for magnetic fields along the different crystal axes.  Fig.~\ref{fig2}b shows the frequency shifts of the most prominent spectral components as a function of applied magnetic field along the crystal b axis.  Because of the symmetry of the Kramer's doublets, the linear Zeeman shift averages to zero for each spectral hole, but it can contribute to broadening of the holes.  The linear Zeeman shift of the most sensitive states is 3.8~kHz/G (magnetic field along the crystal b axis).  To avoid broadening spectral holes at the 1~kHz FWHM level, the magnetic field fluctuations must be less than 0.1~G.

\begin{figure}
\begin{indented}
\item[]\includegraphics[width=0.55\columnwidth]{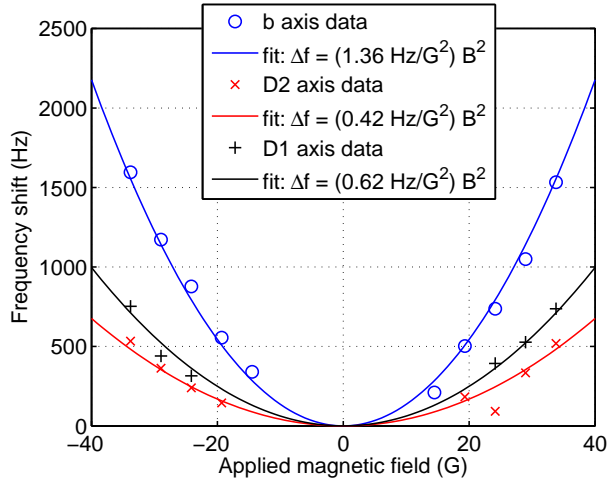}
\end{indented}
\caption{\label{fig3} The quadratic Zeeman shift in 0.1~at.~\%~\EuYSO site 1 spectral holes at 4.5~K for magnetic fields applied along the b-axis, D2-axis, and D1-axis.}
\end{figure}

Unlike the linear Zeeman effect, the quadratic Zeeman effect causes a frequency shift of spectral holes.  Figure~\ref{fig3} shows the area-weighted average (over the most prominent transmission peaks, indicated in Figure~\ref{fig2}a by vertical arrows) frequency shift of a spectral hole written at zero applied field and measured in a nonzero applied magnetic field.  We measure a quadratic frequency sensitivity of 1.4(1.4)~Hz/G$^2$ for magnetic fields along the crystal b axis, which is again the most sensitive direction.  Note that we use the measured value as an estimate of the uncertainty because the atomic population of the original spectral hole is nearly twice the total population of the discernible Zeeman components.  The remaining population is contained in spectral features whose center frequency is not readily discernible.  The measured sensitivity can be compared to the expected value of 0.64 Hz/G$^2$ in free Eu$^{3+}$ atoms~\cite{SobelmanASRT,Dzuba2003}.  In order to reach a fractional frequency instability of $10^{-17}$ the magnetic field must be held at 0~G with an instability better than 60~mG, or in a laboratory field of 0.5~G the magnetic field instability must be below 7~mG.

\section{Thermal effects}\label{sec:tempAndPower}

At low temperature, the frequency shift ($\Delta f$) of spectral holes in \EuYSO exhibits a fourth-order temperature ($T_c$) dependence \cite{Konz2003}.  We measured $\Delta f = \alpha T_c^4$ with $\alpha = 76(15)$~Hz/K$^4$ for site 1 and $\alpha = 250(50)$~Hz/K$^4$ for site 2 over a temperature range of 2.5--5.5~K in a previously described apparatus~\cite{Thorpe2011}.  At 3.65~K, the slope is $15$~kHz/K for site 1 spectral holes and such a high sensitivity could make \EuYSO crystals unsuitable for laser-frequency stabilization.

Here we immerse the SHB crystal in a sealed helium-gas environment to mitigate~\cite{Thorpe2011} the temperature shift~\cite{Konz2003}.  Temperature increases cause the pressure of the helium gas to increase, which compresses the crystal.  Because the spectral-hole frequency shift due to compression is negative and that due to a rise in temperature is positive, the slopes of the two shifts can be made to cancel at a particular pressure of helium gas.  Note, however, that this cancellation works only when the crystal is in thermal equilibrium with the surrounding gas.  In our apparatus a tube connects the sealed sample chamber to a room-temperature gas manifold through which helium can be added and removed.  To reduce time-dependent frequency shifts associated with the thermalization time of the two-reservoir system, we have minimized the volume of the room-temperature portion such that it is approximately equal to the volume of the cryogenic portion.  In the limit $V_r/V_c \ll T_r/T_c$ where $V_r$ ($V_c$) and $T_r$ ($T_c$) are the volume and temperature of the room-temperature (cryogenic) portion of the helium gas, the pressure at which first order temperature shifts cancel is $p_0 = -4 \alpha T_0^4 / \beta$, where $T_0$ is the operating temperature and the spectral-hole pressure shift has been measured \cite{Thorpe2011} to be $\beta = -211.4$~Hz/Pa for site 1 and $\beta = -52.0$~Hz/Pa for site 2.  The residual quadratic temperature sensitivity is $\Delta f = 6 \alpha T_0^2 (T_c - T_0)^2$

\begin{figure}
\begin{indented}
\item[]\includegraphics[width=0.55\columnwidth]{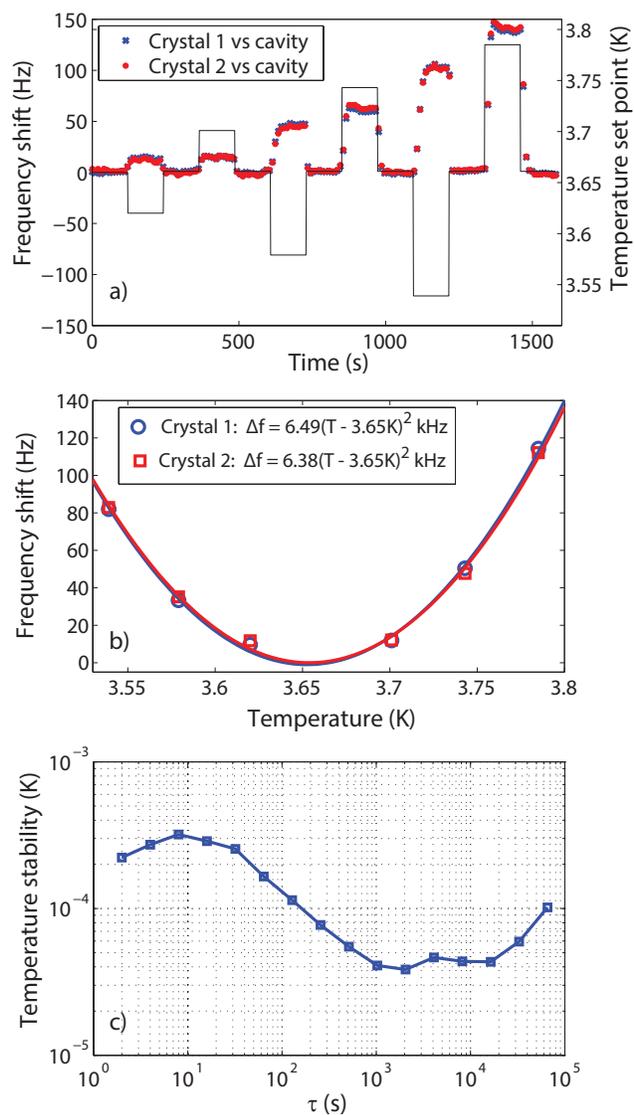}
\end{indented}
\caption{\label{fig4} Calibration of the temperature insensitive point of 1.0~at.~\%~\EuYSO site 1 spectral holes in 270~Pa of helium gas.  (a)  Time-trace of the relative frequency between spectral-hole patterns in crystals 1 and 2 and the pre-stabilization reference cavity.  The temperature is stepped in the vicinity of the insensitive temperature causing shifts in the frequency of the hole pattern. (b)  Fits of the frequency shifts reveal both the insensitive temperature and the quadratic temperature sensitivity. (c)  Temperature stability (Allan deviation) of the sample chamber.  The total measurement duration is $2 \times 10^5$~s.}
\end{figure}

Figures~\ref{fig4}a and \ref{fig4}b show the temperature-dependent frequency shift in 270~Pa of helium gas.  The frequency difference between the pre-stabilization Fabry-P\'{e}rot cavity and a laser that is locked to a pattern of spectral holes is recorded as a function of time while the temperature is changed in discrete steps near the insensitive point.  Figure~\ref{fig4}b shows the frequency shift as a function of the crystal temperature.  The residual quadratic temperature shift at 3.65~K for site 1 spectral holes is 6.5~kHz/K$^2$, which implies that in order to reach $10^{-17}$ fractional frequency instability, the temperature instability needs to be below 1~mK.  If the temperature is 5~mK away from the insensitive temperature, a stability better than 80~$\mu$K is required.

Figure~\ref{fig4}c shows the temperature stability of the sample chamber inside the closed-cycle cryostat whose cold head is thermally stabilized by a feedback loop.  The stability shown here is measured by an out-of-loop sensor located on the sample chamber.  The chamber is thermally decoupled from the cold head by brass washers, which improves the temperature stability of the chamber at short times but degrades it at long times.  Temperature stability is better than 1~mK for the entire range of measurement times and better than 80~$\mu$K for durations between 250~s and 50000~s.

Local heating of the crystal by the laser may cause an additional thermal frequency shift.  However, this effect is reduced because the helium gas inside the chamber thermalizes the crystal with the chamber walls.  We have measured an upper bound for the spectral-hole frequency shift due to a step change in laser power.  We locked two beamlines of the laser separately to a pattern of site 1 spectral holes in each crystal at 2.0~$\mu$W/cm$^2$ laser intensity for 400~s, then increased the laser intensity incident on crystal 2 to 6.6~$\mu$W/cm$^2$ for 50~s (leaving the laser intensity incident on crystal 1 unchanged).  The relative frequency shift was less than 1~Hz, corresponding to a laser intensity sensitivity of less than 0.2~Hz/($\mu$W/cm$^2$).  Current intensity stability is 1~\%, so laser intensity fluctuations are unlikely to limit the performance of this system at the $10^{-17}$ fractional frequency stability level.

\section{Side-holes and anti-holes}\label{sec:sideHoles}

\begin{figure}
\begin{indented}
\item[]\includegraphics[width=0.55\columnwidth]{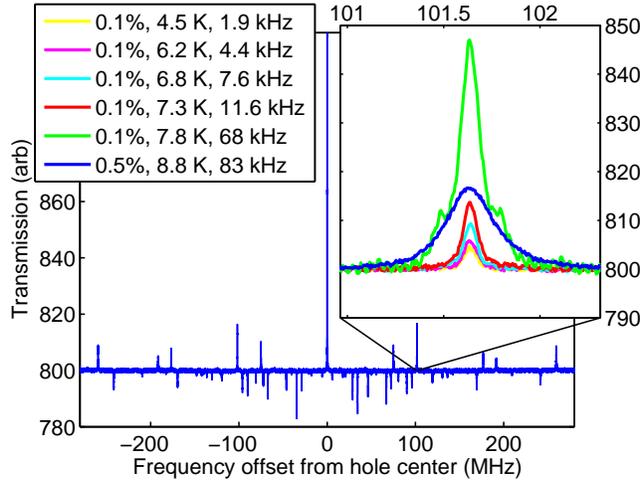}
\end{indented}
\caption{\label{fig5} Side-hole and anti-hole spectrum due to the \EuYSO site 1 spectral-hole hyperfine structure as a function of spectral-hole width and \Eu~doping.  In each transmission measurement, a single spectral hole is burned to a peak amplitude of roughly 600 in the arbitrary transmission units of the figure.  The legend labels the transmission measurements by specifying the \Eu doping level, as well as the crystal temperature and the FWHM of the spectral hole.  The amplitude of the side-holes and anti-holes is reduced by writing narrower spectral holes because the side-holes and anti-holes exhibit inhomogeneous broadening.}
\end{figure}

\begin{figure}
\begin{indented}
\item[]\includegraphics[width=0.55\columnwidth]{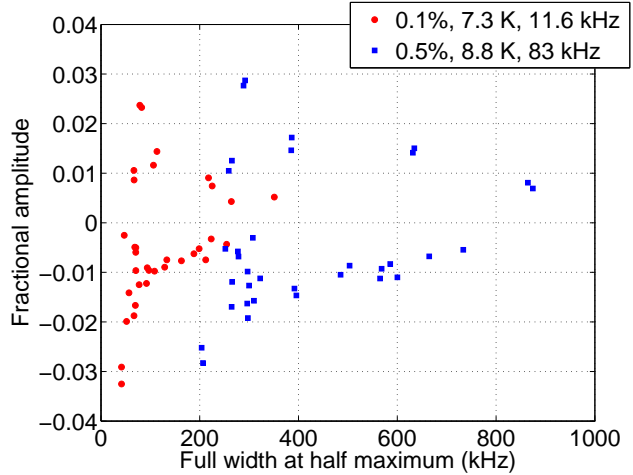}
\end{indented}
\caption{\label{fig6} FWHM and amplitude of the side-holes (positive) and anti-holes (negative) due to the \EuYSO site 1 spectral-hole hyperfine structure as a function of \Eu~doping.  In each transmission measurement, a single 11.6~kHz (0.1\% doping at 7.3~K) or 83~kHz (0.5\% doping at 8.8~K) FWHM spectral hole was burned to full transparency; the transmission amplitude is normalized to the transmission of the main hole.}
\end{figure}

Side-holes and anti-holes \cite{Yano1991} perturb broadband patterns of spectral-holes, which may be the best way to implement a SHB LLO that can operate continuously.  Figure~\ref{fig5} shows the side-hole and anti-hole spectrum of a single spectral hole for several spectral-hole widths and two \Eu doping levels.  Inhomogeneous broadening in the \EuYSO hyperfine structure leads to side-holes and anti-holes that are broader than the main spectral hole feature.  As a result, the amplitude of the side-holes and anti-holes can be reduced by writing narrow spectral holes.  As the \Eu doping level increases, the inhomogeneous broadening in the \EuYSO hyperfine structure increases.  Figure~\ref{fig6} shows the widths and amplitudes of the side-holes and anti-holes of a single spectral hole for two \Eu doping levels.  Increasing the doping from 0.1~\% to 0.5~\% results in a mean increase in the widths of the side-holes and anti-holes by a factor of 3.8.  Perturbations from side-holes and anti-holes to broadband patterns can therefore be reduced by choosing the highest dopant concentration that is compatible with the requirements for spectral-hole lifetime and coherence. 

The effect of side-holes and anti-holes on the performance of a SHB LLO can be estimated by considering how much the center frequency of a spectral hole will be shifted by an overlapping side-hole or anti-hole.  If two 2~kHz FWHM spectral holes of equal depth are burned in 0.5~at.~\%~\EuYSO with worst-case frequency spacing, then side-features of the second hole will shift  the center frequency of the first hole by a fractional-frequency of less than $10^{-17}$ with respect to the optical transition.

\section{Frequency reference shifts}\label{sec:shifts}

\begin{table}
\caption{\label{tab:systematics}Properties and systematic frequency shifts of the \EuYSO crystal.}
\begin{indented}
\item[]\begin{tabular}{@{}lrrl}
\br
Parameter					& \mbox{Site 1}		& \mbox{Site 2}	& Units \\
\mr
Wavelength					& 580.039			& 580.211		& nm \\
Bulk modulus	\cite{Ching2003}		& 134.8			& 134.8		& GPa \\
Volume \cite{Thorpe2011}			& 0.055			& 0.014		& $(\Delta f/f)/(\Delta V/V)$ \\
Magnetic field				& 1.4(1.4)			& -			& Hz/G$^2$ \\
Electric field \cite{Graf1998thesis}	& $< 0.2$			& -			& Hz/(V/m)$^2$ \\
Temperature$^{\rm a}$			& 6.2				& 21			& kHz/K$^2$ \\
Pressure \cite{Thorpe2011}		& -211.4(4)			& -52.0(7)		& Hz/Pa \\
Laser intensity				& $< 0.2$			& -			& Hz/($\mu$W/cm$^2$) \\
Acceleration \cite{Thorpe2011}		& $7 \times 10^{-12}$	& -			& $(\Delta f/f)/g$ \\
\br
\end{tabular}
\item[] $^{\rm a}$ At 3.7~K and 270~Pa (3600~Pa) helium gas pressure for site 1 (2).
\end{indented}
\end{table}

Table~\ref{tab:systematics} summarizes properties of \EuYSO relevant to SHB laser frequency stabilization.  We expect the frequency noise of site 1 spectral holes due to fluctuations in magnetic field, temperature, pressure, and laser intensity to be below $10^{-17}$ in our experimental setup.  Acceleration induced frequency shifts due to vibrations of the closed cycle cryostat may contribute to the spectral hole frequency instability at short averaging time, but will not limit performance at long averaging time.

While the linear Stark coefficient of \EuYSO site 1 spectral holes has been measured \cite{Graf1998thesis} to be 350(50)~Hz/(V/m), the more relevant quadratic Stark effect has not yet been measured.  Similar to the Zeeman effect, when the amplitude of the electric field is small, the linear Stark effect causes spectral-hole broadening and the quadratic Stark effect causes the center frequency of the spectral holes to shift.  The linear Stark effect data of Graf \cite{Graf1998thesis} are consistent with a zero quadratic Stark effect with a large uncertainty of $\pm0.2$~Hz/(V/m)$^2$ for \EuYSO site 1 spectral holes.

The rate of spectral-hole frequency drift due to background electric and magnetic field drift can be estimated using the linear and quadratic field sensitivities together with a measurement of the rate of spectral-hole broadening.  We have measured the rate of \EuYSO site 1 spectral-hole broadening in the absence of broadening due to continuous probing by burning a 2.1~kHz FWHM spectral hole and measuring the width every 30~minutes.  The observed growth rate of the spectral-hole FWHM, 0.04~Hz/s, indicates that the rate of change of background fields is likely below $6 \times 10^{-5}$~(V/m)/s and $5 \times 10^{-6}$~G/s.  In a laboratory magnetic field of 0.5~G, such a magnetic field drift would cause less than $7 \times 10^{-6}$~Hz/s of spectral-hole frequency drift.

\section{Conclusion}

Spectral-hole burning in \EuYSO shows much promise for improved LLOs.  The systematic frequency shifts of \EuYSO spectral holes due to magnetic field, temperature, helium gas pressure, and optical probe power fluctuations are all small enough to allow laser frequency stabilization at the $10^{-17}$ level in the current apparatus. 

We have shown qualitatively that perturbations to the spectral-hole pattern due to side-holes and anti-holes can be reduced by using narrower spectral holes and higher doping levels, but more work is required to determine the contribution of side-holes and anti-holes to SHB LLO frequency noise.  Further studies of the quadratic Stark shift and the thermomechanical noise floor are also needed, but we are optimistic that spectral-hole burning can reduce the frequency noise of lasers and extend their coherence time.

\section*{Acknowledgments}

We thank R.~L.~Cone, J.~C.~Bergquist and D.~J.~Wineland for useful discussions, and S.~Cook and C.~W.~Oates for critical readings of this manuscript.  This work is supported by DARPA QuASAR and ONR, and is not subject to US copyright.

\section*{References}

\bibliographystyle{iopart-num}

\begin{thebibliography}{10}
\expandafter\ifx\csname url\endcsname\relax
  \def\url#1{{\tt #1}}\fi
\expandafter\ifx\csname urlprefix\endcsname\relax\def\urlprefix{URL }\fi
\providecommand{\eprint}[2][]{\url{#2}}

\bibitem{Chou2010a}
Chou C~W, Hume D~B, Koelemeij J~C~J, Wineland D~J and Rosenband T 2010 {\em
  Phys. Rev. Lett.\/} {\bf 104} 070802

\bibitem{Jiang2011}
Jiang Y~Y, Ludlow A~D, Lemke N~D, Fox R~W, Sherman J~A, Ma L~S and Oates C~W
  2011 {\em Nature Photonics\/} {\bf 5} 158--61

\bibitem{Kessler2012}
Kessler T, Hagemann C, Grebing C, Legero T, Sterr U, Riehle F, Martin M~J, Chen
  L and Ye J 2012 {\em Nature Photonics\/} {\bf 6} 687--92

\bibitem{Numata2004}
Numata K, Kemery A and Camp J 2004 {\em Phys. Rev. Lett.\/} {\bf 93} 250602

\bibitem{Notcutt2006}
Notcutt M, Ma L~S, Ludlow A~D, Foreman S~M, Ye J and Hall J~L 2006 {\em Phys.
  Rev. A\/} {\bf 73} 031804

\bibitem{Rosenband2008}
Rosenband T, Hume D~B, Schmidt P~O, Chou C~W, Brusch A, Lorini L, Oskay W~H,
  Drullinger R~E, Fortier T~M, Stalnaker J~E, Diddams S~A, Swann W~C, Newbury
  N~R, Itano W~M, Wineland D~J and Bergquist J~C 2008 {\em Science\/} {\bf 319}
  1808--12

\bibitem{Blatt2008}
Blatt S, Ludlow A~D, Campbell G~K, Thomsen J~W, Zelevinsky T, Boyd M~M, Ye J,
  Baillard X, Fouch\'{e} M, {Le Targat} R, Brusch A, Lemonde P, Takamoto M,
  Hong F~L, Katori H and Flambaum V~V 2008 {\em Phys. Rev. Lett.\/} {\bf 100}
  140801

\bibitem{Chou2010b}
Chou C~W, Hume D~B, Rosenband T and Wineland D~J 2010 {\em Science\/} {\bf 329}
  1630--3

\bibitem{Zhang2010}
Zhang W, Xu Z, Lours M, Boudot R, Kersalé Y, Santarelli G and {Le Coq} Y 2010
  {\em Appl. Phys. Lett.\/} {\bf 96} 211105

\bibitem{Fortier2011}
Fortier T~M, Kirchner M~S, Quinlan F, Taylor J, Bergquist J~C, Rosenband T,
  Lemke N, Ludlow A, Jiang Y, Oates C~W and Diddams S~A 2011 {\em Nature
  Photonics\/} {\bf 5} 425--9

\bibitem{Benedick2012}
Benedick A~J, Fujimoto J~G and K\"{a}rtner F~X 2012 {\em Nature Photonics\/}
  {\bf 6} 97--100

\bibitem{Sellin1999a}
Sellin P~B, Strickland N~M, Carlsten J~L and Cone R~L 1999 {\em Opt. Lett.\/}
  {\bf 24} 1038--40

\bibitem{Strickland2000}
Strickland N~M, Sellin P~B, Sun Y, Carlsten J~L and Cone R~L 2000 {\em Phys.
  Rev. B\/} {\bf 62} 1473--6

\bibitem{Bottger2003}
B\"{o}ttger T, Pryde G~J and Cone R~L 2003 {\em Opt. Lett.\/} {\bf 28} 200--2

\bibitem{Julsgaard2007}
Julsgaard B, Walther A, Kr\"{o}ll S and Rippe L 2007 {\em Opt. Express\/} {\bf
  15} 11444--65

\bibitem{Thorpe2011}
Thorpe M~J, Rippe L, Fortier T~M, Kirchner M~S and Rosenband T 2011 {\em Nature
  Photonics\/} {\bf 5} 688--93

\bibitem{Konz2003}
K\"{o}nz F, Sun Y, Thiel C~W, Cone R~L, Equall R~W, Hutcheson R~L and
  Macfarlane R~M 2003 {\em Phys. Rev. B\/} {\bf 68} 085109

\bibitem{Yano1991}
Yano R, Mitsunaga M and Uesugi N 1991 {\em Opt. Lett.\/} {\bf 16} 1884--86

\bibitem{Macfarlane1994}
Equall R~W, Sun Y, Cone R~L and Macfarlane R~M 1994 {\em Phys. Rev. Lett.\/}
  {\bf 72} 2179--2182

\bibitem{Chen2011}
Chen Q~F, Troshyn A, Ernsting I, Kayser S, Vasilyev S, Nevsky A and Schiller S
  2011 {\em Phys. Rev. Lett.\/} {\bf 107} 223202

\bibitem{Drever1983}
Drever R~W~P, Hall J~L, Kowalski F~V, Hough J, Ford G~M, Munley A~J and Ward H
  1983 {\em Appl. Phys. B\/} {\bf 31} 97--105

\bibitem{Leibrandt2011a}
Leibrandt D~R, Thorpe M~J, Notcutt M, Drullinger R~E, Rosenband T and Bergquist
  J~C 2011 {\em Opt. Express\/} {\bf 19} 3471--82

\bibitem{Li1992}
Li C, Wyon C and Moncorg\'{e} R 1992 {\em IEEE J. Quant. Electron.\/} {\bf 28}
  1209--21

\bibitem{Nazarova2006}
Nazarova T, Riehle F and Sterr U 2006 {\em Appl. Phys. B\/} {\bf 83} 531--6

\bibitem{Shelby1981}
Shelby R~M and Macfarlane R~M 1981 {\em Phys. Rev. Lett.\/} {\bf 47} 1172--5

\bibitem{SobelmanASRT}
Sobelman I~I 1979 {\em {Atomic Spectra and Radiative Transitions}\/}

\bibitem{Dzuba2003}
Dzuba V~A, Safronova U~I and Johnson W~R 2003 {\em Phys. Rev. A\/} {\bf 68}
  032503

\bibitem{Ching2003}
Ching W, Ouyang L and Xu Y~N 2003 {\em Phys. Rev. B\/} {\bf 67} 245108

\bibitem{Graf1998thesis}
Graf F~R 1998 {\em {Investigations of spectral dynamics in rare earth ion doped
  crystals using high resolution laser techniques}\/} Ph.D. thesis ETH
  Z\"{u}rich

\end{thebibliography}
\providecommand{\newblock}{}

\end{document}